\documentclass[reprint,tightenlines,showpacs,twocolumn,longbibliography,aps,prd]{revtex4-2}
\usepackage[pagewise]{lineno}
\usepackage{graphicx}
\usepackage{dcolumn}
\usepackage{bm}
\usepackage{rotating}
\usepackage{amsmath}
\usepackage{amssymb}
\usepackage{hyperref}
\hypersetup{colorlinks=true,linkcolor=blue,anchorcolor=blue,citecolor=blue,urlcolor=blue}
\usepackage{overpic}
\usepackage{booktabs}
\usepackage{threeparttable}
\usepackage{float}
\usepackage{makecell}
\usepackage{multirow}

\newcommand{\BR}{{\mathcal{B}}}

\newcommand{\psip}{\psi(3686)}
\newcommand{\psipp}{\psi(3770)}

\newcommand{\jpsi}{J/\psi}

\newcommand{\EE}{e^+e^-}
\newcommand{\LLb}{\Lambda\bar{\Lambda}}

\newcommand{\pp}{\pi^+\pi^-}

\newcommand{\bfg}{\begin{figure}}
\newcommand{\efg}{\end{figure}}
\newcommand{\bitm}{\begin{itemize}}
\newcommand{\eitm}{\end{itemize}}
\newcommand{\bnum}{\begin{enumerate}}
\newcommand{\enum}{\end{enumerate}}
\newcommand{\btbl}{\begin{table}}
\newcommand{\etbl}{\end{table}}
\newcommand{\btbu}{\begin{tabular}}
\newcommand{\etbu}{\end{tabular}}

\newcommand{\beq}{\begin{equation}}
\newcommand{\edq}{\end{equation}}

\begin{document}
\normalsize
\parskip=5pt plus 1pt minus 1pt

\title{\boldmath Measurement of $ \psi (3686)\to \LLb \eta$ and $ \psi (3686)\to \LLb \pi^0 $ decays}
\author
{
    \begin{small}
        \begin{center}
            M.~Ablikim$^{1}$, M.~N.~Achasov$^{11,b}$, P.~Adlarson$^{70}$, M.~Albrecht$^{4}$, R.~Aliberti$^{31}$, A.~Amoroso$^{69A,69C}$, M.~R.~An$^{35}$, Q.~An$^{66,53}$, X.~H.~Bai$^{61}$, Y.~Bai$^{52}$
% ...
            \\
            \vspace{0.2cm}
            (BESIII Collaboration)\\
            \vspace{0.2cm} {\it
                $^{1}$ Institute of High Energy Physics, Beijing 100049, People's Republic of China\\
% ...
$^{4}$ Bochum Ruhr-University, D-44780 Bochum, Germany\\
% ...
$^{11}$ G.I. Budker Institute of Nuclear Physics SB RAS (BINP), Novosibirsk 630090, Russia\\
% ...
$^{31}$ Johannes Gutenberg University of Mainz, Johann-Joachim-Becher-Weg 45, D-55099 Mainz, Germany\\
% ...
$^{35}$ Liaoning Normal University, Dalian 116029, People's Republic of China\\
% ...
$^{52}$ Southeast University, Nanjing 211100, People's Republic of China\\
                $^{53}$ State Key Laboratory of Particle Detection and Electronics, Beijing 100049, Hefei 230026, People's Republic of China\\
% ...
$^{61}$ University of Jinan, Jinan 250022, People's Republic of China\\
% ...
$^{66}$ University of Science and Technology of China, Hefei 230026, People's Republic of China\\
% ...
$^{69}$ University of Turin and INFN, (A)University of Turin, I-10125, Turin, Italy; (B)University of Eastern Piedmont, I-15121, Alessandria, Italy; (C)INFN, I-10125, Turin, Italy\\
                $^{70}$ Uppsala University, Box 516, SE-75120 Uppsala, Sweden\\
% ...
$^{b}$ Also at the Novosibirsk State University, Novosibirsk, 630090, Russia\\
% ...
            }\end{center}
        \vspace{0.4cm}
    \end{small}
}
\date{\today}

\begin{abstract}

    Based on a sample of $448.1\times10^6\ \psip$ events collected with the BESIII detector, a study of $\psip\to\LLb\pi^0$
    and $\psip\to\LLb\eta$ is performed. Evidence of the isospin-violating decay $\psip\to\LLb\pi^0$ is found for the first time
    with a statistical significance of $3.7\sigma$, the branching fraction $\mathcal{B}(\psip\to\LLb\pi^0)$ is measured to be
    $(1.42\pm0.39\pm0.59)\times10^{-6}$, and its corresponding upper limit is determined to be $2.47\times 10^{-6}$ at 90\%
    confidence level. A partial wave analysis of $\psip\to\LLb\eta$ shows that the peak around $\Lambda\eta$ invariant mass threshold
    favors a $\Lambda$ resonance with mass and width in agreement with the $\Lambda(1670)$. The branching fraction of the
    $\psip\to\LLb\eta$ is measured to be $(2.34\pm0.18\pm0.52)\times10^{-5}$. The first uncertainties are statistical and the second are
    systematic.

\end{abstract}


\maketitle

{\centering\section{Introduction}}

The $\psip$ is the first radial excitation of the isospin singlet $c\bar{c}$ vector state, and its decays involving baryon pairs
not only provide an opportunity to study the baryon structure, but also offer a unique place to investigate SU(3) flavor symmetry
breaking effects. Using a sample of 107 million $\psip$ events collected in 2009, $\mathrm{BESIII}$ published a study investigating the
reactions of $ \psip \to \LLb \pi ^0$ and $ \psip \to \LLb \eta$~\cite{BESIII:2012jve}. However, no significant signal for
$ \psip \to \LLb \pi ^0$ was seen, due to suppression from isospin conservation, and an upper limit of
$ \BR ( \psip \to \LLb \pi ^0)<2.9 \times 10^{-6}$ was reported at the 90\% confidence level (CL). Based on roughly $60$ events,
the first branching fraction measurement of $ \psip \to \LLb \eta $ was obtained:
$ \BR ( \psip \to \LLb \eta )=(2.47 \pm 0.34 \pm 0.19) \times 10^{-5}$.

In 2012, another sample of $\psip$ events was collected at the BESIII detector~\cite{BESIII:2017tvm}. The total data set of 448 million
$\psip$ events, corresponding to a four-fold increase of 2009 data, allows for an in-depth investigation on the decays
$ \psip \to \LLb \pi ^0$ and $ \psip \to \LLb \eta$, and searches for intermediate $\Lambda$ states in the $\Lambda\pi^0$ and
$\Lambda\eta$ mass spectra. In addition, these branching fractions may also be used to test the ``12\%'' rule
\cite{Duncan:1980qd,Brodsky:1981kj,Chernyak:1981zz}, which predicts that the ratio of branching fractions of $\psip$ and $\jpsi$ decays
into the same light hadron final states is around 12\%. In this paper, the charge-conjugate process is always implied unless explicitly
mentioned.\\

{\centering\section{Detector and Monte Carlo samples}}

The BESIII detector~\cite{BESIII:2009fln} records symmetric $e^+e^-$ collisions provided by the BEPCII storage ring~\cite{Yu:2016cof},
in the center-of-mass energy range from 2.0~GeV to 4.95~GeV, with a peak luminosity of $1\times10^{33}\;\mathrm{cm}^{-2}\mathrm{s}^{-1}$
achieved at $\sqrt{s} = 3.77\;\mathrm{GeV}$.
BESIII has collected large data samples in this energy region~\cite{BESIII:2020nme}. The cylindrical core of the BESIII detector
covers 93\% of the full solid angle and consists of a helium-based multilayer drift chamber (MDC), a plastic scintillator
time-of-flight system~(TOF), and a CsI(Tl) electromagnetic calorimeter (EMC), which are all enclosed in a superconducting solenoidal
magnet providing a 1.0~T magnetic field. The solenoid is supported by an octagonal flux-return yoke equipped with resistive plate counter
muon identification modules interleaved with steel. The charged-particle momentum resolution at $1~{\rm GeV}/c$ is $0.5\%$, and the
resolution of the specific energy loss $dE/dx$ in the MDC is $6\%$ for electrons from Bhabha scattering. The EMC measures photon
energies with a resolution of $2.5\%$ ($5\%$) at $1$~GeV in the barrel (end cap) region. The time resolution in the TOF barrel region
is 68~ps, while that in the end cap region is 110~ps.

Simulated data samples produced with a GEANT4-based~\cite{GEANT4:2002zbu} Monte Carlo (MC) package, which includes the geometric
description of the BESIII detector and the detector response, are used to optimize the event selection criteria, determine detection
efficiencies and estimate backgrounds. The simulation models the beam energy spread and initial state radiation (ISR) in $e^+e^-$
annihilations with the generator {\sc kkmc}~\cite{Jadach:2000ir,Jadach:1999vf}. The inclusive MC sample aims to include all
possible processes involving the production of the $ \jpsi $ and $ \psip $ resonances, and the continuum processes incorporated in
    {\sc kkmc} \cite{Jadach:2000ir,Jadach:1999vf}. The known decay modes are modeled with {\sc evtgen}~\cite{Lange:2001uf,Ping:2008zz} using
branching fractions taken from the Particle Data Group (PDG)~\cite{Workman:2022ynf}, and the remaining unknown charmonium
decays are modeled with {\sc lundcharm} \cite{Chen:2000tv,Yang:2014vra}. Final state radiation (FSR) from charged particles is incorporated using
    {\sc photos}~\cite{Richter-Was:1992hxq}. Signal MC samples of $ \psip \to \LLb \pi ^0$ decays are generated with uniform phase space
(PHSP), while $ \psip \to \LLb \eta $ decays are generated according to the results of the partial wave analysis reported later
in this paper. \\

{\centering\section{Event Selection}}
The processes $\psip\to\LLb\pi^0$ and $\psip\to\LLb\eta$ are reconstructed with $\Lambda\rightarrow~p\pi^{-}$, $\bar{\Lambda}\rightarrow~\bar{p}\pi^{+}$, $\pi^{0}\rightarrow~\gamma\gamma$ and $\eta\to\gamma\gamma$. Since the final state for both channels is $p\bar{p}\pi^+\pi^-\gamma\gamma$, the number of charged tracks is required to be four with net charge zero. Each track
must satisfy $\left|\cos\theta\right|<0.93$, where $\theta$ is the polar angle of the track measured by the MDC with respect
to the direction of the positron beam.

Each of the photon candidates is required to have an energy deposit in the EMC of at least 25 MeV in the barrel
$( \left| \cos \theta \right| <0.80)$ or 50 MeV in the end caps $(0.86< \left| \cos \theta \right| <0.92)$. To eliminate showers
from charged tracks, the angle between the position of each shower in the EMC and any charged track must be greater than 10 degrees.
To suppress electronic noise and showers unrelated to the event, the EMC time difference from the event start time is required to be
within $[0,700]$ ns. At least two photon candidates are required.

\begin{figure*}[htbp]
    \centering
    \begin{overpic}[width=0.48\textwidth]{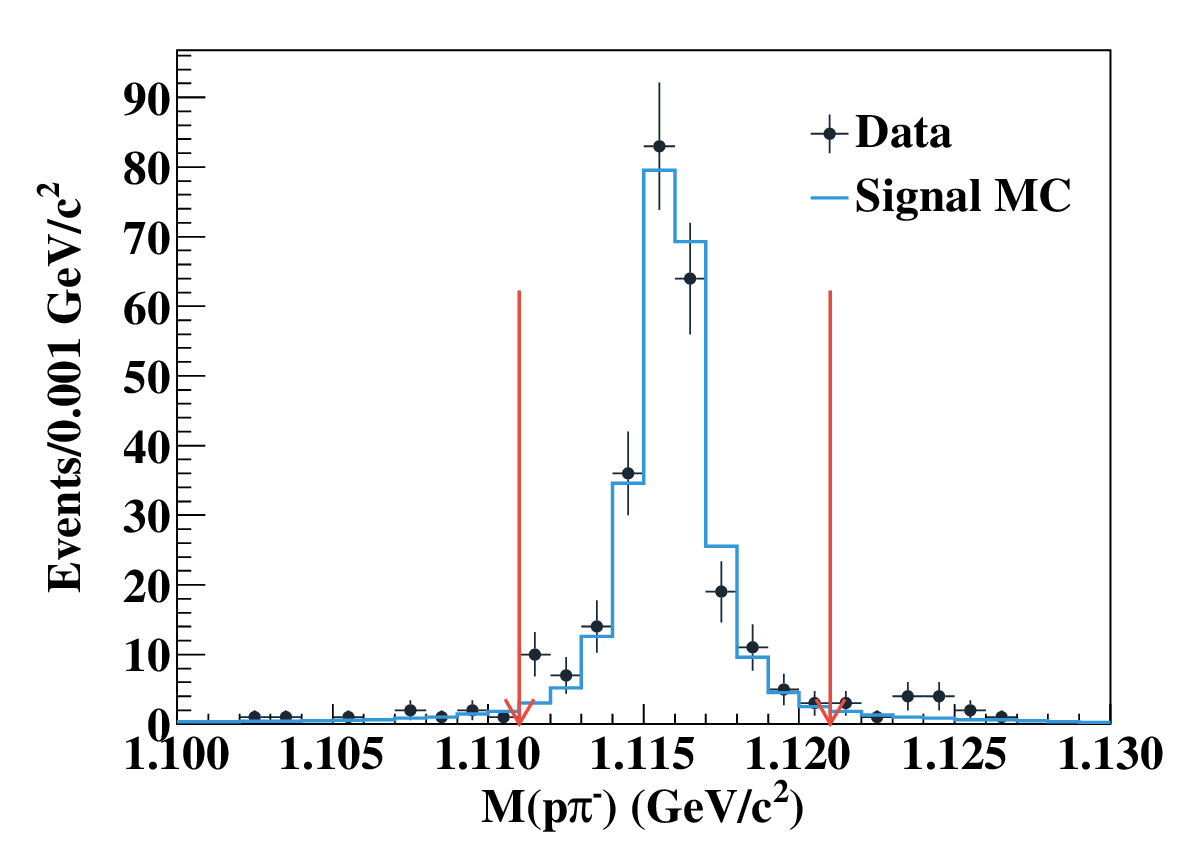}
        \put(22,57){$(a)$}
    \end{overpic}
    \begin{overpic}[width=0.48\textwidth]{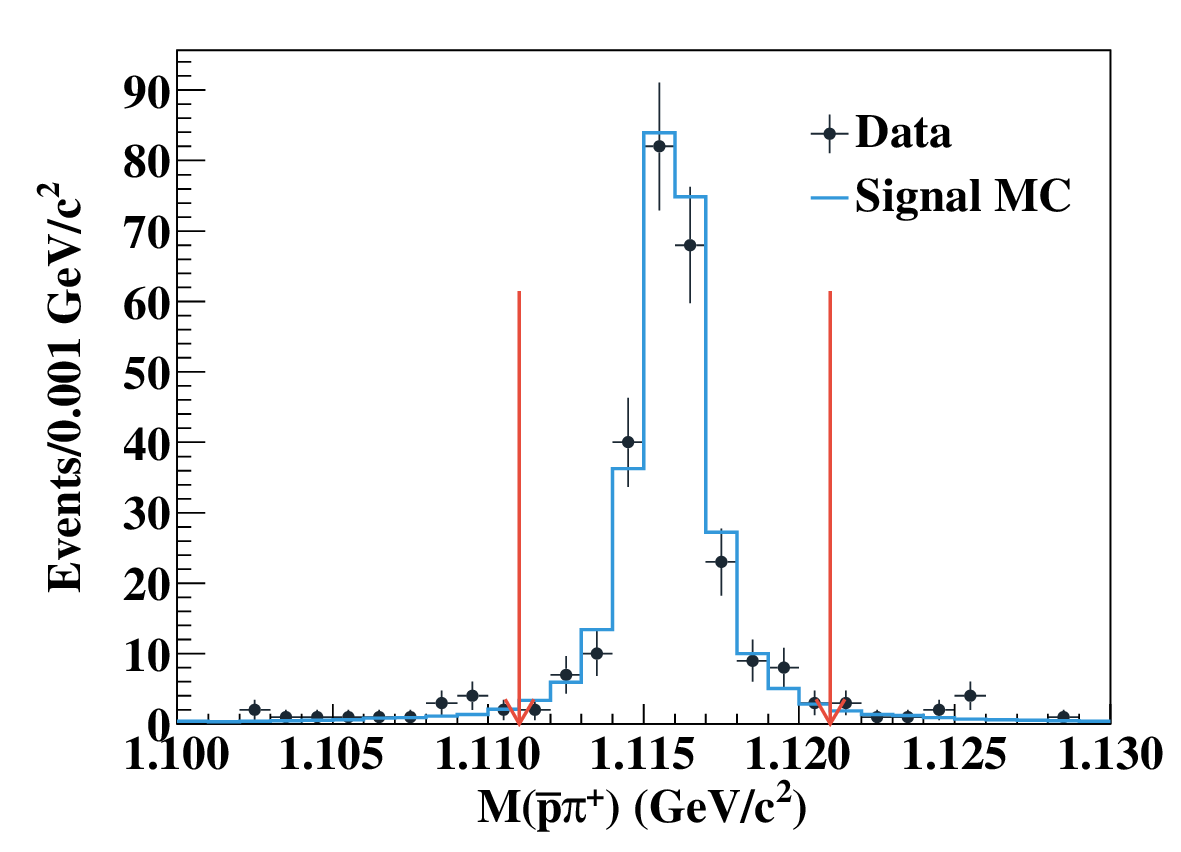}
        \put(22,57){$(b)$}
    \end{overpic}
    \caption{The distributions of (a) $ \mathrm{M} (p \pi ^-)$ and (b) $ \mathrm{M} ( \bar{p} \pi ^+)$.
        Dots with error bars represent data, the blue histograms are normalized signal MC.
        The mass window requirement of the $\Lambda(\bar{\Lambda})$ is shown with the red arrows.}
    \label{fig:line}
\end{figure*}

The $ \Lambda $ and $ \bar{ \Lambda } $ candidates are reconstructed by combining pairs of oppositely charged tracks with pion and
proton mass hypotheses, fulfilling a secondary vertex constraint~\cite{Xu:2009zzg}.
Events with at least one $p\pi^-(\Lambda)$ and one $\bar{p}\pi^+(\bar{ \Lambda })$ candidate are selected.
In the case of multiple $ \LLb $ pair candidates, the one with the minimum
value of $ \chi ^2_{svtx}( \Lambda )+ \chi ^2_{svtx}( \bar{ \Lambda } )$ is chosen, where $ \chi ^2_{svtx}( \Lambda )$ and
$ \chi ^2_{svtx}( \bar{ \Lambda } )$ are the fit qualities of the secondary vertex fits for $ \Lambda $ and $ \bar{ \Lambda } $,
respectively. To improve the momentum and energy resolution and to reduce background contributions, a four-constraint (4C)
energy-momentum conservation kinematic fit is applied to the event candidates under the hypothesis of $\LLb \gamma\gamma$ (i.e., not
considering the $\gamma\gamma$ mass), and the corresponding $\chi^2_{4C}$ is required to be less than 40. For events with more than two
photon candidates, the combination with the best fit quality is selected from all possible combinations. To reject possible background
contributions from $\psip\to \LLb \gamma$ and $\psip\to \LLb \gamma\gamma\gamma$, we further require that the $\chi^2$ of the 4C fit for
the $\psip \to \LLb \gamma\gamma$ assignment is smaller than those of $ \LLb \gamma $ and $ \LLb \gamma \gamma \gamma $. The final
$p \pi ^-$ and $ \bar{p} \pi ^+$ mass distributions in two progresses are shown in Figs.~\ref{fig:line} (a) and (b), respectively,
where clear $\Lambda$ and $\bar{\Lambda}$ signals are visible.

We require that the invariant mass of $p \pi ^-$($ \bar{p} \pi ^+$) should be in the mass region of $\Lambda$,
$1.111<\mathrm{M} (p \pi ^-, \bar{p} \pi ^+) < 1.121\ \mathrm{GeV}/c^2$. To remove the background events from $\psip\to\pi^0(\eta)\jpsi$,
$\jpsi\to\LLb$, events with the invariant mass of $\LLb$ in the $J/\psi$ mass region, $3.087<M(\LLb)<3.107$
GeV/$c^2$, are rejected. In order to suppress possible background events from $\psip\to \pi^+\pi^- J/\psi$ with
$J/\psi\to p\bar{p}\pi^0(\eta)$, we reject events with $3.087<\mathrm{M} _{rec}( \pp )<3.107\ \mathrm{GeV}/c^2$, where
$ \mathrm{M} _{rec}( \pp )$ is the mass recoiling against the $\pi^+\pi^-$.
No requirements on the $\gamma\gamma$ invariant mass are imposed since this is the variable which will be used to extract the signal yields.

In the case of $ \psip \to \LLb \pi ^0$, additional requirements are applied to further reduce the contamination from background. The $\chi^2_{4C}$
is required to be less than 15, to further suppress background events with one or more than two photons in the final states. The veto cut
$\mathrm{M}(\LLb)<3.4\ \mathrm{GeV}/c^2$ is applied in order to suppress background from $ \psip \to \Sigma ^0 \bar{ \Sigma } ^0$.
Two other veto cuts, $\mathrm{M}(p\pi^0, \bar{p}\pi^0)<1.17\ \mathrm{GeV}/c^2$ and $\mathrm{M}(p\pi^0, \bar{p}\pi^0)>1.2\ \mathrm{GeV}/c^2$,
are applied in order to suppress background from $\psip\to\Lambda\bar{\Sigma}^-\pi^+$. The invariant masses $\mathrm{M}(\gamma_{low}\Lambda)$
and $\mathrm{M}(\gamma_{low}\bar{\Lambda})$ are both required to be outside of $(1.183,1.203)\ \mathrm{GeV}/c^2$ to suppress the
$ \psip \to \Lambda \bar{ \Sigma } ^0 \pi ^0$ background, where $\gamma_{low}$ represents the less energetic candidate photon.

    {\centering\section{Background study}\label{sec:4}}

To investigate the possible background contributions, the same selection criteria are applied to an inclusive MC sample
of 506 million $\psip$ events. A topological  analysis of the surviving events is performed with the generic tool TopoAna
\cite{Zhou:2020ksj}, and the results indicate that the background peaking at the $\pi^0$ invariant mass mainly comes from
$\psip \to \Lambda \bar{ \Sigma } ^0 \pi ^0$, while the other background sources present a flat distribution. Thus, a PHSP MC sample of
$ \psip \to \Lambda \bar{ \Sigma } ^0 \pi ^0 + c.c.$ is generated, giving a background estimate of $20.4 \pm 1.9$ events,
by using a branching fraction of $(1.54 \pm 0.04 \pm 0.13) \times 10^{-4}$ obtained from $\mathcal{B}(\psip\to\Lambda\bar{\Sigma}^-\pi^+)$
with isospin symmetry considerations.

To estimate the background from $e^+e^-$ continuum processes, the same procedure is performed on data taken at
$ \sqrt{s} = 3.773\ \mathrm{GeV} $, with an integrated luminosity of $2.92 \ \mathrm{fb} ^{-1}$~\cite{Ablikim:2013ntc}.
The background events are extracted by fitting the $M_{\gamma\gamma}$ mass distribution, normalized to the $\psip$ data
taking into account the luminosity and energy-dependent cross section of the quantum electrodynamics (QED) processes. The normalization
factor $f$ is calculated as

\begin{equation*}
    f=\frac{N_{\psip}}{N_{\psipp}}=\frac{\mathcal{L}_{\psip}}{\mathcal{L}_{\psipp}}\cdot\frac{\sigma_{\psip}}{\sigma_{\psipp}}\cdot\frac{\epsilon_{\psip}}{\epsilon_{\psipp}},
\end{equation*}

\noindent where $N$, $ \mathcal{L} $, $ \sigma $, and $ \epsilon $ refer to the number of observed events, integrated luminosity of data,
cross section, and detection efficiency at the two center of mass energies, respectively. The details on the cross section values can be found in
Ref.~\cite{Asner:2008nq}. The detection efficiency ratio $\epsilon_{\psip} / \epsilon_{\psipp}$ has been
determined by Monte Carlo simulations. After normalization, the background contributions from $ \EE \to \LLb \pi ^0$ and
$ \EE \to \LLb \eta $ at $3.686 \ \mathrm{GeV} $ are determined to be $13.2 \pm 1.7$ and $19.1 \pm 2.0$ events, respectively.

Due to the identical event topology, these background events are indistinguishable from signal events and are subtracted directly
by fixing their magnitudes in the fit when extracting signal yields. Here we assume that the interference between $\psip$ decay and
continuum process is negligible. \\

{\centering\section{\boldmath Analysis of $\psip\to \Lambda\bar{\Lambda} \pi ^0$}}

The $ \psip \to \LLb \pi ^0$ signal yield is obtained from an extended unbinned maximum likelihood fit to the $\gamma\gamma$ invariant mass distribution.
The total probability density function consists of a signal and various background contributions. The signal component is modeled
with the MC simulated signal shape convolved with a Gaussian function to account for a possible difference in the mass resolution
between data and MC simulation. The background events from $e^+e^-$ annihilations are described by the shape obtained from the
data taken at $ \sqrt{s} =3.773\ \mathrm{GeV}$, while the peaking background from $\psip \to \Lambda \bar{ \Sigma } ^0 \pi ^0$
is modeled with the MC simulation shape. In the fit, the contributions of these two background sources are fixed to the values
discussed above. In addition, the nonpeaking background is parameterized by a first order Chebychev function.

\begin{figure}[htbp]
    \centering
    \includegraphics[width=0.49\textwidth]{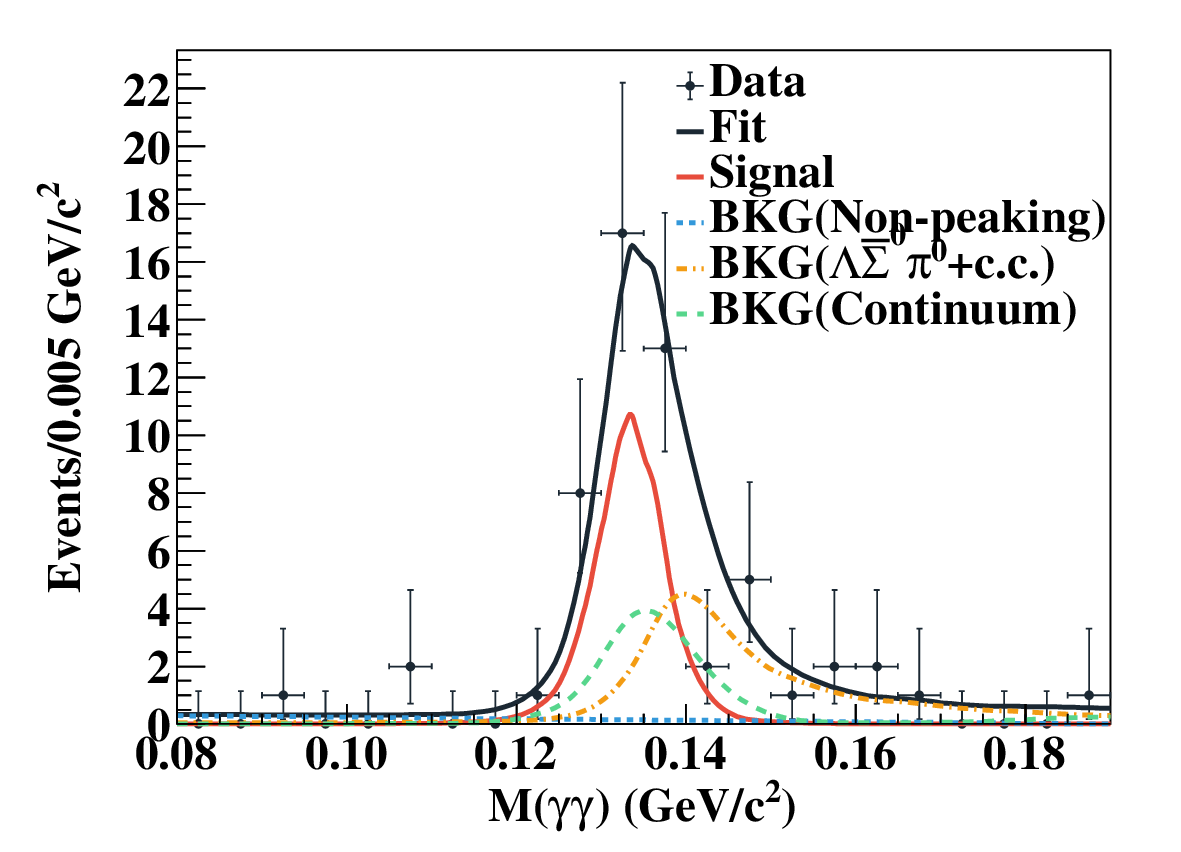}
    \caption{The distribution of $ \mathrm{M}( \gamma \gamma )$ in the $\pi^0$ region. Dots with error bars are data, the black solid curve is
        the fit result, the red solid curve represents the signal, the blue dashed curve is the nonpeaking background, the orange dash-dotted solid curve
        is $ \psip \to \Lambda \bar{ \Sigma } ^0 \pi ^0$, and the green long-dashed curve is the continuum background.}
    \label{fig:fit_m_pi0}
\end{figure}

From the fit, shown in Fig.~\ref{fig:fit_m_pi0}, we estimate $23.0 \pm 6.3$ $ \Lambda\bar{\Lambda} \pi ^0$ events with a statistical
significance of $3.7\sigma$ which is evaluated by comparing the likelihood values with and without the $\pi^0$ signal included in the fit. The
detection efficiency obtained from MC simulation events is 9.0\% and these results are summarized in Table~\ref{tab:value_bf_cal}. \\

{\centering\section{\boldmath Analysis of $ \psip \to \LLb \eta $}}

\begin{figure}[htbp]
    \centering
    \includegraphics[width=0.49\textwidth]{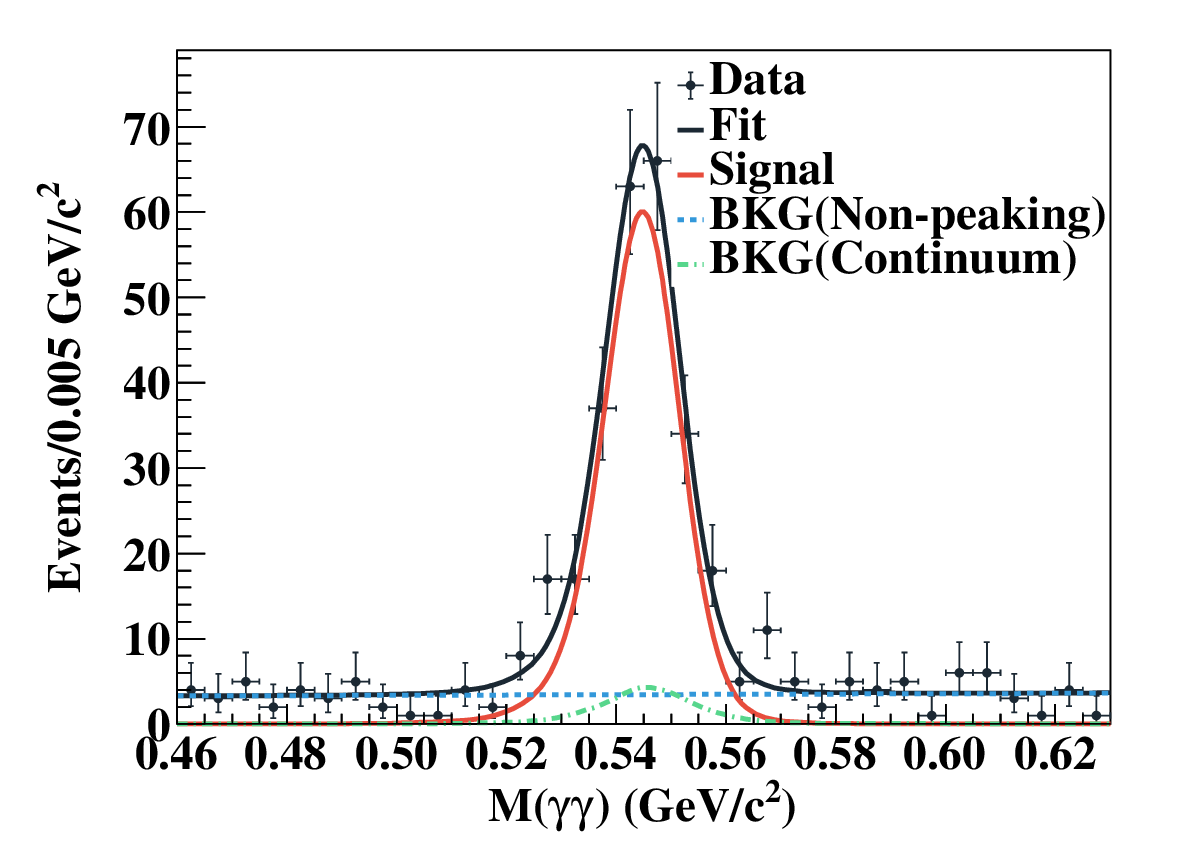}
    \caption{The distribution of $ \mathrm{M}( \gamma \gamma )$ in the $\eta$ mass region. Dots with error bars are data, the black solid curve is the fit result, the red solid
        curve represents the signal, the blue dashed curve is the nonpeaking background, and the green dash-dotted curve is the continuum background.}
    \label{fig:fit_m_eta}
\end{figure}

The distribution of $M(\gamma\gamma)$ in the $\eta$ mass region is shown in Fig.~\ref{fig:fit_m_eta}. A fit to the $\eta$ signal with the MC
simulated signal shape convolved with a Gaussian function is performed, and the background contribution is described by the
shape obtained from the continuum data plus a first order Chebychev function. The fitting results are shown in Fig.
\ref{fig:fit_m_eta}, with a total of $218 \pm 17$ $ \Lambda\bar{\Lambda} \eta$ signal events.

\begin{figure}[htbp]
    \centering
    \includegraphics[width=0.48\textwidth]{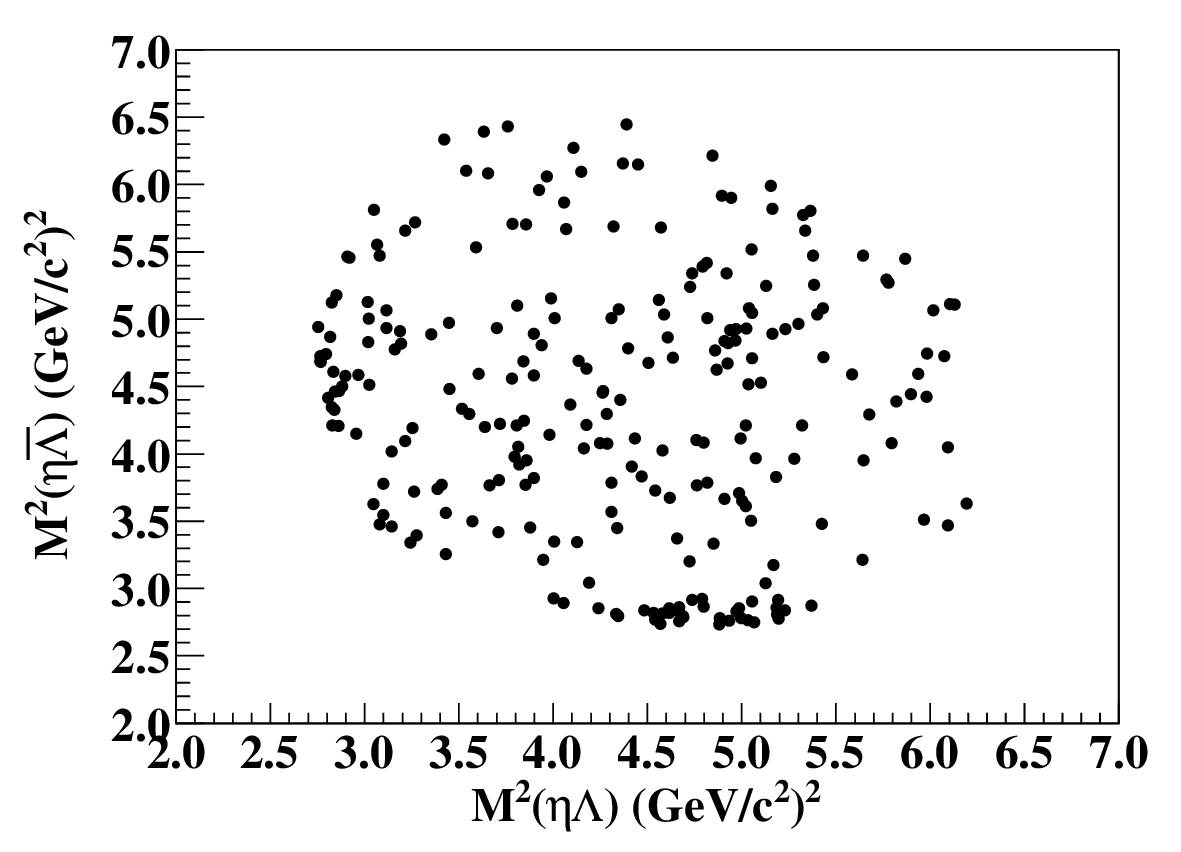}
    \caption{The Dalitz plot of $\mathrm{M}^2(\eta\Lambda)$ versus $\mathrm{M}^2(\eta\bar{\Lambda})$.}
    \label{fig:eta_pwa_dz}
\end{figure}

\begin{figure*}[htbp]
    \centering
    \begin{overpic}[width=0.32\textwidth]{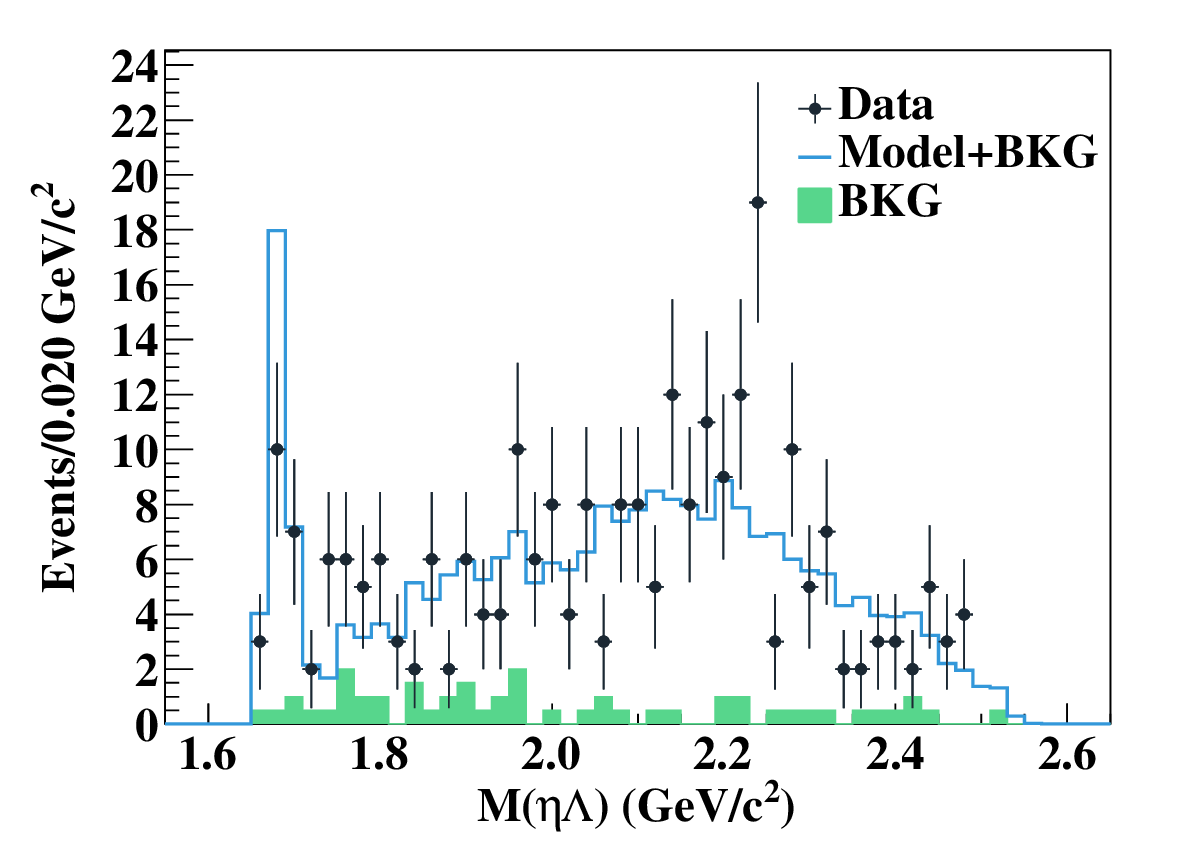}
        \put(35,57){$(a)$}
    \end{overpic}
    \begin{overpic}[width=0.32\textwidth]{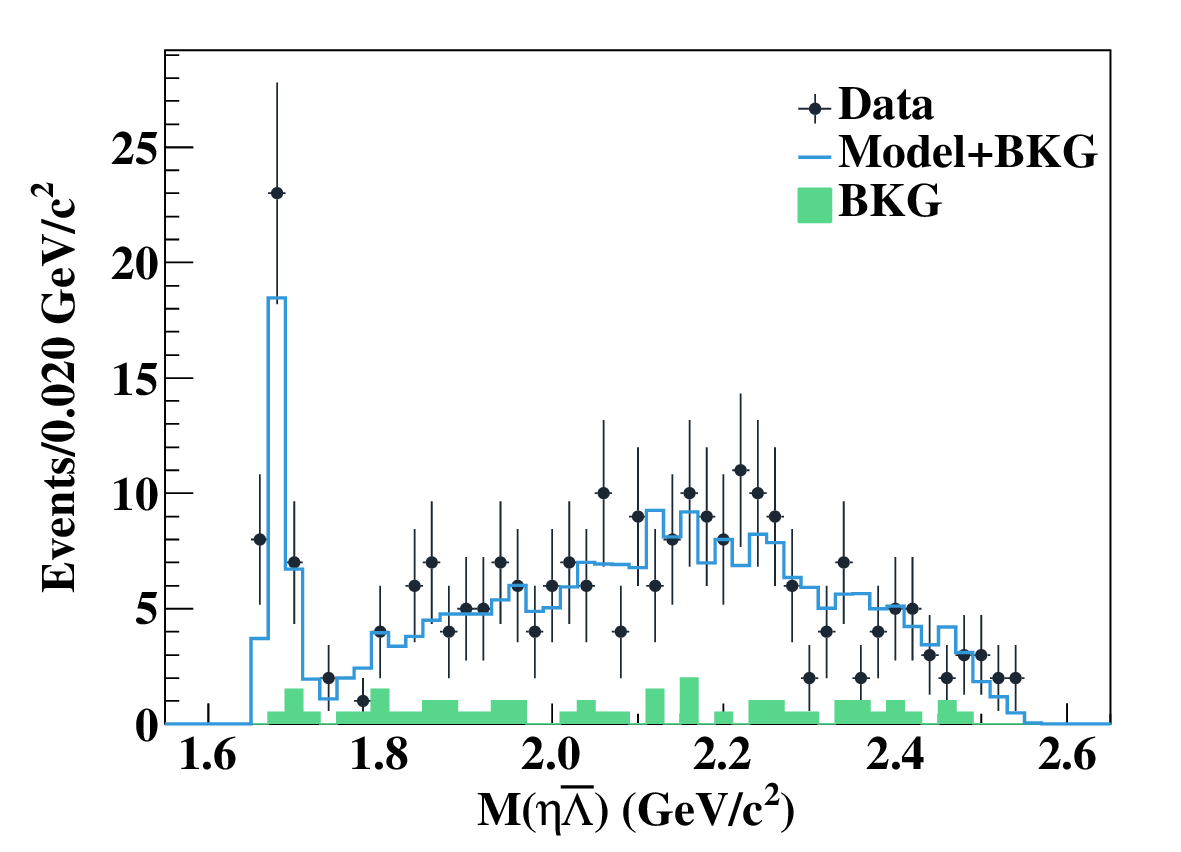}
        \put(35,57){$(b)$}
    \end{overpic}
    \begin{overpic}[width=0.32\textwidth]{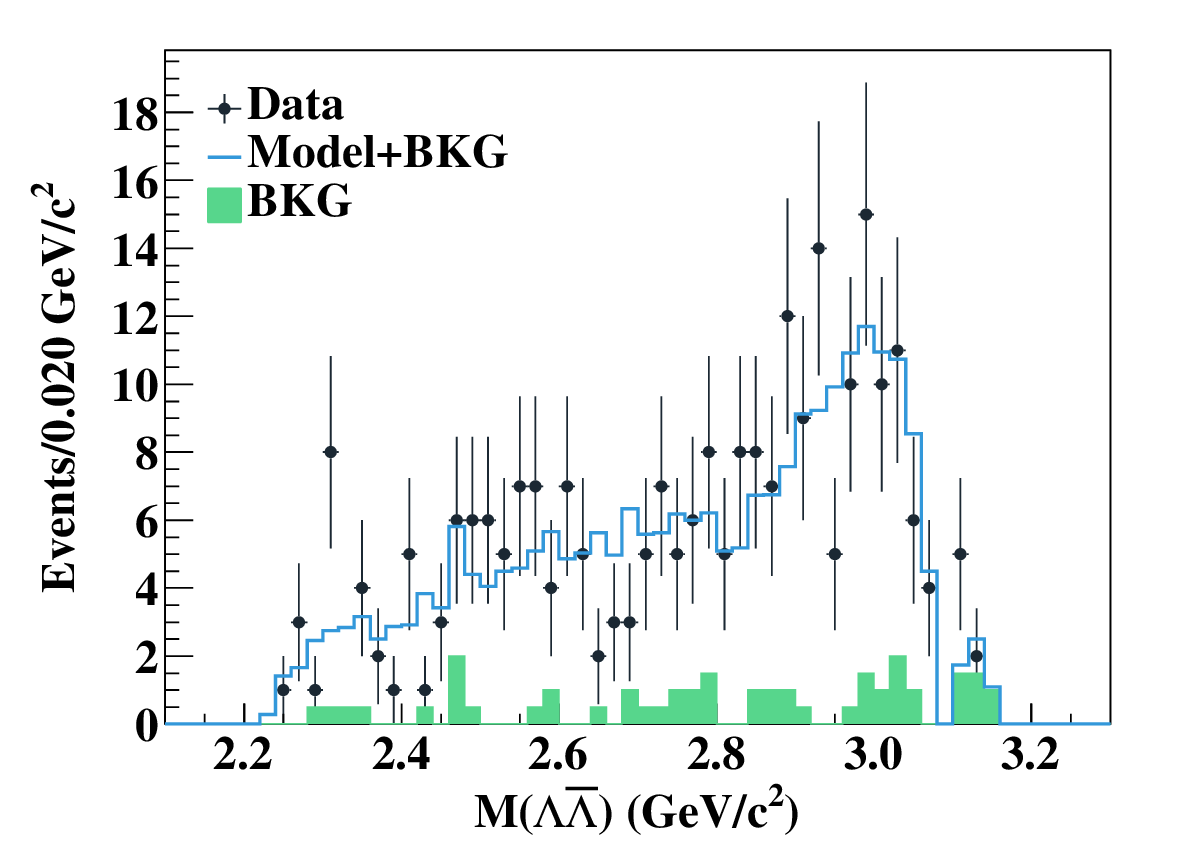}
        \put(55,57){$(c)$}
    \end{overpic}
    \caption{The distributions of (a) $ \mathrm{M} ( \eta \Lambda )$, (b) $ \mathrm{M} ( \eta \bar{ \Lambda } )$ and (c)
        $ \mathrm{M} ( \Lambda \bar{ \Lambda } )$. Dots with error bars represent data, the blue histograms are the sum of the PWA
        results and $\eta$ sidebands, and the background contributions estimated from the $ \eta $ sidebands are indicated with the
        green shaded histograms.}
    \label{fig:eta_pwa_m}
\end{figure*}

To investigate possible intermediate states, $\mathrm{M}(\gamma\gamma)$ is required to be in the $\eta$ range $(0.525,0.560)\ \mathrm{GeV}/c^2$. The
resultant Dalitz plot of the 252 selected $\psip\to\LLb\eta$ candidates, shown in Fig.~\ref{fig:eta_pwa_dz}, exhibits two visible clusters
of events around $2.8\ (\mathrm{GeV}/c^2)^2$ in $\mathrm{M}^2(\eta\Lambda)$ and $\mathrm{M}^2(\eta\bar{\Lambda})$, indicating possible
intermediate excited baryons. The invariant mass distributions of $\mathrm{M}(\eta\Lambda)$ and $\mathrm{M}(\eta\bar{\Lambda})$ presented in
Figs.~\ref{fig:eta_pwa_m} (a) and (b) show the corresponding structures. The $\eta$ sidebands,
$0.470\ \mathrm{GeV}/c^2<\mathrm{M}(\gamma\gamma)<0.505\ \mathrm{GeV}/c^2$ and
$0.580\ \mathrm{GeV}/c^2<\mathrm{M}(\gamma\gamma)<0.615\ \mathrm{GeV}/c^2$, are used to estimate the number of background events;
the obtained distributions, shown as shaded histograms in Figs.~\ref{fig:eta_pwa_m} (a), (b) and (c), indicate that the structures
are not from background events.

Using the Feynman diagram calculation package~\cite{Wang:2004du}, a partial wave analysis (PWA) is performed based on an unbinned maximum likelihood fit.
In the global fit, resonances are described by a relativistic Breit-Wigner propagator, with the mass and width as free parameters,

\begin{equation}
    \mathrm{BW}(s)=\frac{1}{M^2_{\Lambda^*}-s-iM_{\Lambda^*}\Gamma_{\Lambda^*}},
\end{equation}

\noindent where $s$ is the squared invariant mass.

\begin{figure*}[htbp]
    \centering
    \begin{overpic}[width=0.32\textwidth]{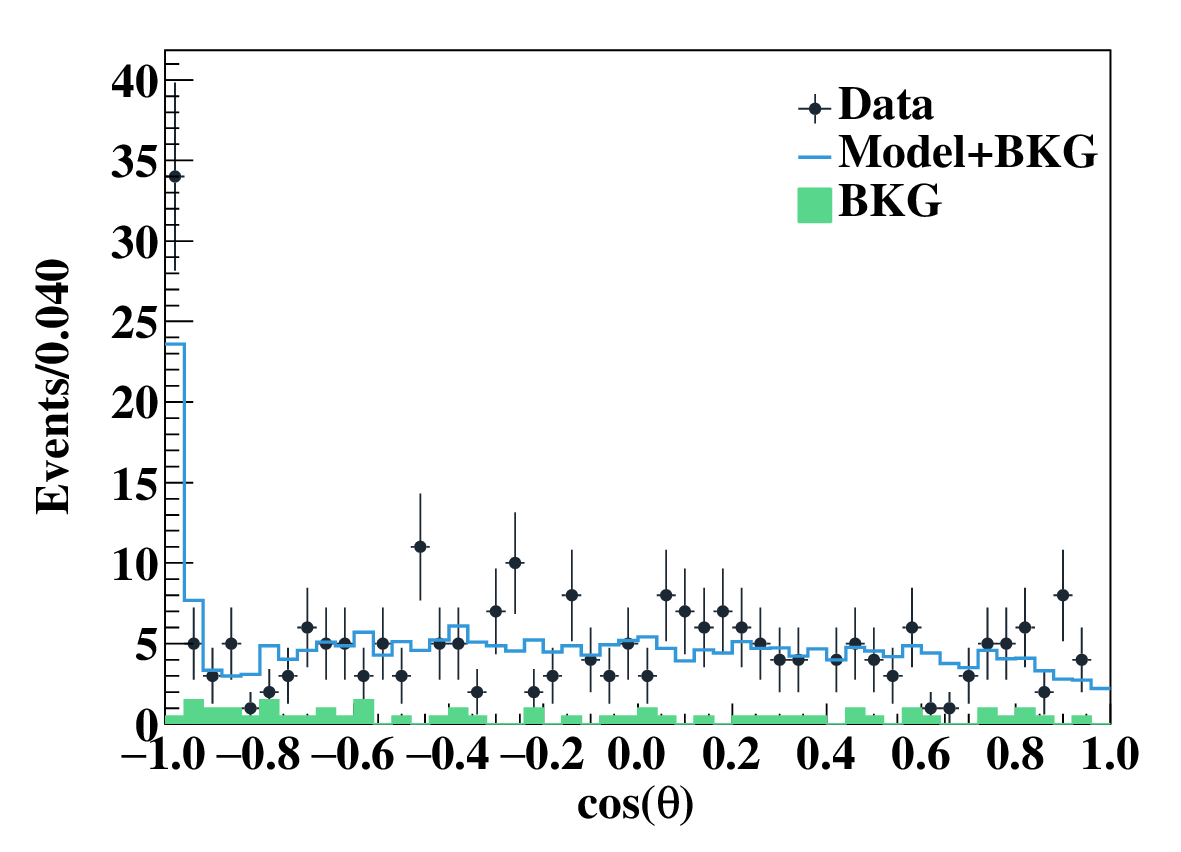}
        \put(35,57){$(a)$}
    \end{overpic}
    \begin{overpic}[width=0.32\textwidth]{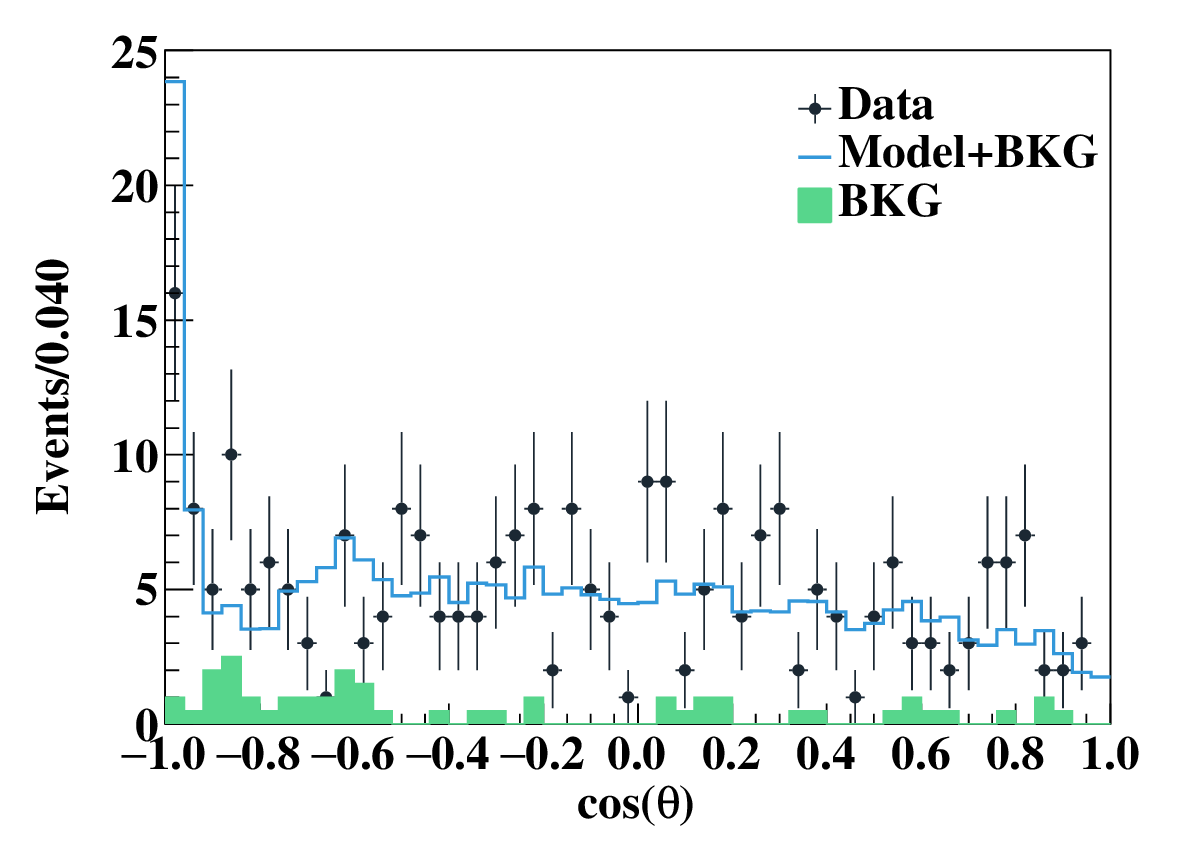}
        \put(35,57){$(b)$}
    \end{overpic}
    \begin{overpic}[width=0.32\textwidth]{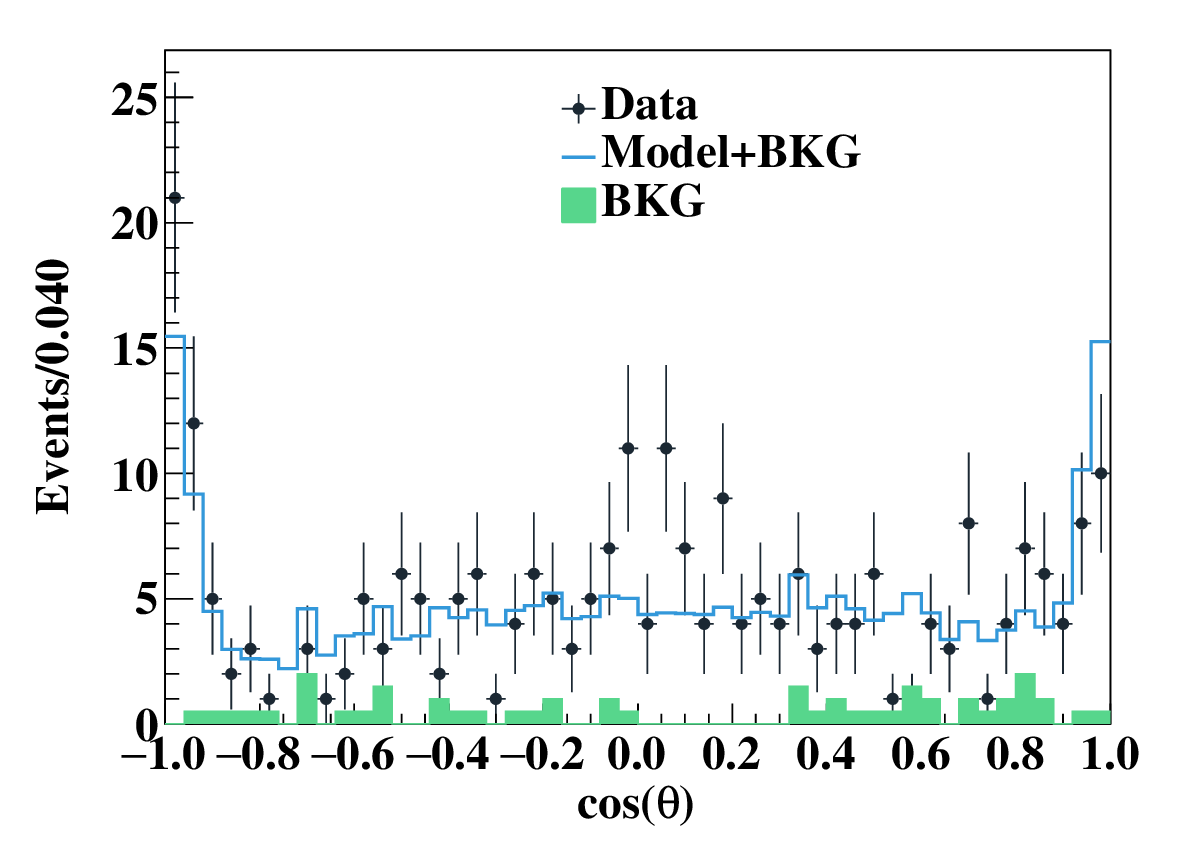}
        \put(25,57){$(c)$}
    \end{overpic}
    \caption{The distributions of (a) $ \cos \theta $ between $ \Lambda $ and $ \bar{ \Lambda } $ in the center-of-mass system
        (CMS) of $\eta \Lambda$, (b) $ \cos \theta $ between $ \Lambda $ and $ \bar{ \Lambda } $ in CMS of $\eta \bar{ \Lambda }$
        and (c) $ \cos \theta $ between $ \eta $ and $ \Lambda $ in CMS of $\Lambda \bar{ \Lambda }$. The dots with error bars
        represent data, the blue histograms are the sum of the PWA results and $\eta$ sidebands, and the green shaded histograms are
        the background events from $\eta$ sidebands.}
    \label{fig:eta_pwa_cos}
\end{figure*}

To describe the $\Lambda\eta$ and $\bar{\Lambda}\eta$ mass spectra, all kinematically-allowed resonances of $\Lambda^*$ and $\Sigma^*$
listed in the PDG~\cite{Workman:2022ynf} are considered. Only components with a statistical significance larger
than $5\sigma$ are kept in the baseline solution. PWA results indicate that the $\Lambda(1670)$ plus the nonresonant contribution could
provide a good description of data, as illustrated in Figs.~\ref{fig:eta_pwa_m} and~\ref{fig:eta_pwa_cos}. The fitted mass and width of $\Lambda(1670)$,
$(1672 \pm 5) \ \mathrm{MeV}/c^2$ and $(38\pm 10)$ MeV, are also in agreement with the world average values;
a total of $116 \pm 28 \ \psip \to \Lambda(1670)\bar{\Lambda} $ candidate events are measured (based on the PWA amplitude ``fit fraction''), and the detection efficiency is determined to be
12.5\% by using a PWA-weighted MC sample. The measured yield and detection efficiency are summarized
in Table~\ref{tab:value_bf_cal}.
The hypothesis of a $\Lambda(1690)$ state instead of $\Lambda(1670)$ in the model has been tested, leading to a reasonable description of the data, but with a sightly worse fit quality and with resonance parameters not consistent with the PDG values; it has thus been rejected.

To obtain the detection efficiency of $\psip \to \Lambda\bar{\Lambda} \eta$, a MC sample is generated in accordance with
the above PWA results. The corrected detection efficiency, 12.9\%, and the number of signal events, $218 \pm17$ are presented in
Table~\ref{tab:value_bf_cal}.\\

{\centering\section{Systematic uncertainties}}

In this analysis, the systematic uncertainties on the branching fractions mainly come from the following sources:

\begin{itemize}

    \item $\Lambda$ reconstruction

          The efficiency of $ \Lambda ( \bar{ \Lambda } )$ reconstruction is studied using the control sample of
          $ \psip \to \Lambda \bar{ \Lambda } $ decays, and a correction factor of $0.980 \pm 0.011$~\cite{BESIII:2017kqw}
          is applied to the efficiencies obtained from MC simulation. The uncertainty of the correction factor, 1.1\%,
          which includes the uncertainties of MDC tracking and $ \Lambda ( \bar{ \Lambda } )$ reconstruction, is considered as
          the uncertainty of the efficiency of $ \Lambda ( \bar{ \Lambda } )$ reconstruction.

    \item Photon detection

          The photon detection efficiency has been studied using a high-purity control sample of $\jpsi\to\rho^0\pi^0$
          \cite{BESIII:2011ysp}. The difference between the detection efficiencies of data and MC is
          around 1\% per photon. Thus, 2\% is assigned as the total systematic uncertainty for the detection of the two photons.

    \item Kinematic fit

          The uncertainty associated with the 4C kinematic fit comes from the inconsistency between data and MC simulation in the fit.
          This difference is reduced by correcting the track helix parameters of the MC simulation, with parameters from
          \cite{BESIII:2013nam} and~\cite{BESIII:2019gjc}. Following the method described in Ref.~\cite{BESIII:2012mpj}, we obtain
          the systematic uncertainties of the 4C kinematic fit as 3.8\% and 1.8\% for $\psip\to\Lambda\bar{\Lambda}\pi^0$ and
          $\psip\to\Lambda\bar{\Lambda}\eta$, respectively.

    \item Mass window requirements

          The systematic uncertainties related to each individual mass window requirement are estimated by varying the size of the mass
          window by one standard deviation of the corresponding mass resolution. For the mass window of
          $\mathrm{M}(\LLb)<3.4\ \mathrm{GeV}/c^2$, the uncertainty is estimated by decreasing the required mass threshold by
          $10\ \mathrm{MeV}/c^2$. The largest variation of branching fraction for each mass requirement is considered as the related systematic uncertainty.

    \item Signal shape

          In order to estimate the systematic uncertainty due to the signal shape, alternative fits are performed to determine the
          yields of signal events; the MC shape is replaced with a Breit-Wigner function convolved with a Gaussian function or a single
          Gaussian function, by varying the fits of the invariant mass distributions by either contracting, expanding or shifting the fit range
          by $\pm10\ \mathrm{MeV}$. The maximum differences with the nominal results are assigned as the corresponding systematic
          uncertainties.

    \item Background uncertainty

          To estimate the uncertainty of the nonpeaking background shape in the fit to $ \mathrm{M} ( \gamma \gamma )$, we performed
          alternative fits by replacing the first-order Chebychev function with a second-order Chebychev function for $ \psip $
          data. The maximum changes of 2.0\% and 3.5\% are considered as systematic uncertainties.
          The uncertainties of background from continuum events and the decay $\psip\to\bar{\Lambda}\Sigma^0\pi^0$ are propagated
          from the statistical uncertainties quoted in Sec.~\ref{sec:4}.

    \item Interference between $\psip$ and continuum amplitudes

          To estimate the effect from interference of the continuum amplitude with the resonance amplitude, we use the method from Ref.~\cite{Guo:2022gkg}. The maximum impact from interference term with respect to the resonance term is defined as $r_R^{max}$,

          \begin{equation}
              r_R^{max}=\frac{4}{\hbar c}AB,\ A=\sqrt{\frac{\sigma_c^f(s)}{\mathcal{B}_f}},
          \end{equation}

          where $\hbar c$ is the conversion constant, $\sigma_c^f(s)$ is the cross section of the continuum process measured from data,
          $\mathcal{B}_f$ is the branching fraction of $\psip\to\Lambda\bar{\Lambda}\pi^0$ and $\psip\to\Lambda\bar{\Lambda}\eta$ that
          we measured in this paper and the factor $B$ is constant depending on the resonance parameters quoted from
          Ref.~\cite{Guo:2022gkg}. The $r_R^{max}$, $40.3\%$ and $20.6\%$ of $\psip\to\Lambda\bar{\Lambda}\pi^0$ and
          $\psip\to\Lambda\bar{\Lambda}\eta$, are taken as the uncertainty of interference, respectively. Since the
          $\Lambda(1670)\bar{\Lambda}$ cannot be studied well in continuum with our current statistics, no systematic is provided for
          this final state.

    \item Physics model

          To have a good description of data from $ \psip \to \Lambda \bar{ \Lambda } \eta $, an event generator based on the
          PWA results is developed to determine the detection efficiency. We vary the default configuration to a setup either
          with only the $\Lambda(1690)$ or with a combination of $\Lambda (1670)$ and $\Lambda(1690)$, and consider the largest change
          in the detection efficiency as systematic uncertainty of the physical model. For $ \psip \to \Lambda \bar{ \Lambda } \pi ^0$,
          we use PHSP as the nominal event generator. The change in detection efficiency using an alternative model within the allowed
          phase space of the $\Lambda\pi^0$ system resonances is assigned as systematic uncertainty for the $\LLb\pi^0$ model.

    \item Intermediate decays

          The uncertainties of the quoted decay branching fractions for the intermediate particles from PDG~\cite{Workman:2022ynf}
          are taken as systematic uncertainties.

    \item Number of $\psip$ events

          The number of $ \psip $ events is determined from an analysis of inclusive hadronic $ \psip $ decays. The uncertainty of the
          number of $ \psip $ events, 0.6\%~\cite{BESIII:2017tvm}, is taken as systematic uncertainty.
\end{itemize}

\begin{table*}[htbp]
    \centering
    \caption{ The systematic uncertainties for the (product) branching
        fractions of $ \psip \to \Lambda \bar{ \Lambda } \pi^0 $, $ \psip \to \Lambda \bar{ \Lambda } \eta $ and
        $ \psip \to \Lambda \bar{ \Lambda } (1670)$. All values are given in percent.}
    \label{tab:eta_syu}
    \begin{threeparttable}
        \begin{tabular}{l|c|c|c}
            \hline
            \hline
            Source                                    & $ \psip \to \Lambda \bar{ \Lambda } \pi^0 $ & $ \psip \to \Lambda \bar{ \Lambda } \eta $ & $ \psip \to \Lambda \bar{ \Lambda } (1670)$ \\
            \hline
            $ \Lambda(\bar{\Lambda})$ reconstruction  & 1.1\tnote{b}                                & 1.1                                        & 1.1                                         \\
            Photon detection                          & 2.0\tnote{b}                                & 2.0                                        & 2.0                                         \\
            Kinematic fit                             & 3.8\tnote{b}                                & 1.8                                        & 1.8                                         \\
            Mass window requirements                  & 7.1\tnote{b}                                & 5.1                                        & 5.1                                         \\
            Signal shape                              & 0.9\tnote{a}                                & 1.2                                        & -                                           \\
            Nonpeaking background                     & 2.0\tnote{a}                                & 3.5                                        & -                                           \\
            Peaking background                        & 7.3\tnote{a}                                & 1.0                                        & -                                           \\
            \makecell[c]{Interference between $\psip$                                                                                                                                          \\ and continuum amplitudes} & 40.3\tnote{a}                            & 20.6                                       & -                                           \\
            Physics model                             & 1.4\tnote{b}                                & 3.4                                        & -                                           \\
            $ \BR ( \Lambda \to p \pi )$              & 0.8\tnote{b}                                & 0.8                                        & 0.8                                         \\
            $ \BR ( \pi^0 (\eta) \to \gamma \gamma )$ & 0.03\tnote{b}                               & 0.5                                        & 0.5                                         \\
            Number of $ \psip $ events                & 0.6\tnote{b}                                & 0.6                                        & 0.6                                         \\
            PWA additional resonances                 & -                                           & -                                          & 34.2                                        \\
            PWA background                            & -                                           & -                                          & 12.9                                        \\
            PWA PHSP parametrization                  & -                                           & -                                          & 30.6                                        \\
            \hline
            Total                                     & 41.9                                        & 22.1                                       & 48.0                                        \\
            \hline
        \end{tabular}
        \begin{tablenotes}
            \footnotesize
            \item[a] Additive
            \item[b] Multiplicative
        \end{tablenotes}
    \end{threeparttable}
\end{table*}

In addition, the systematic uncertainties associated with the PWA, which contribute to the measurement of the $\Lambda(1670)$ mass and width
and of the corresponding production branching fraction, are described below.

\begin{itemize}

    \item Additional resonances

          To investigate the impact on the PWA results from other possible components, the analysis has been performed including
          additional possible states (e.g. $\Lambda(1690)$). The changes of the mass, width, and fitted fraction of $\Lambda(1670)$
          are considered as systematic uncertainties, and the largest one was chosen.

    \item Background uncertainty

          In the $\psip\to\LLb\eta$ decays, the background level is quite low, and the events from the $ \eta $ sidebands are considered in the PWA.
          To estimate the uncertainty, the scale factor of background events from $ \eta $ sidebands has been varied by $\pm50\%$, and the
          largest variation of the results is assigned as systematic uncertainty.

    \item PHSP parametrization

          In the partial wave analysis, PHSP is parametrized as a resonance with an extremely large width, and fixed values of spin and parity. The contribution to the
          systematic uncertainty is estimated by replacing
          the spin parity of $\frac{1}{2}^-$ with $\frac{1}{2}^+$, $\frac{3}{2}^-$ or $\frac{3}{2}^+$. The largest resulting
          difference is taken as systematic uncertainty.
\end{itemize}

All the systematic uncertainty sources and values are summarized in Tables~\ref{tab:eta_syu} and~\ref{tab:eta_pwa_syu}, respectively, for the $\psip\to\Lambda\bar{\Lambda}\pi^0$, $\psip\to\Lambda\bar{\Lambda}\eta$ decays, where the total uncertainties are given by the quadratic sum, assuming statistical independence of all the contributions. The distinction between additive and multiplicative sources of systematic uncertainties are indicated in the table.

\begin{table}[htbp]
    \centering
    \caption{Summary of the systematic uncertainty sources contributing to the mass ($\Delta \mathrm{M}$) and width ($\Delta \mathrm{\Gamma}$) of $\Lambda(1670)$.}
    \label{tab:eta_pwa_syu}
    \resizebox*{0.49\textwidth}{!}{\begin{tabular}{lcc}
            \hline
            \hline
            Source                    & $ \Delta \mathrm{M} ( \mathrm{MeV}/c^2 )$ & $ \Delta \Gamma ( \mathrm{MeV} )$ \\
            \hline
            PWA additional resonances & 6                                         & 12                                \\
            PWA background            & 2                                         & 2                                 \\
            PWA PHSP parametrization  & 1                                         & 14                                \\
            \hline
            Total                     & 6                                         & 19                                \\
            \hline
        \end{tabular}}
\end{table}

{\centering\section{Results}}

The branching fractions of the decays of interest are calculated as

\begin{align}
     & \BR ( \psip \to \Lambda \bar{ \Lambda } X) = \notag                                                                      \\
     & = \frac{N^{obs}_{X}}{N_{ \psip } \cdot \BR ^2( \Lambda \to p \pi ^-) \cdot \BR ( X \to \gamma \gamma ) \cdot \epsilon },
\end{align}

\noindent where $X$ is $\pi^0$ or $\eta$, $N^{obs}_{X}$ is the number of signal candidates, $N_{ \psip }$ is the number of $ \psip $
events determined with inclusive hadronic events, $\epsilon$ is the detection efficiency obtained from the MC simulation.
$ \BR ( \Lambda \to p \pi ^-)$, $ \BR ( \pi ^0 \to \gamma \gamma )$ and $ \BR ( \eta \to \gamma \gamma )$ are the corresponding
branching fractions from PDG~\cite{Workman:2022ynf}. Using the numbers given in Table~\ref{tab:value_bf_cal}, the branching
fractions of $\psip \to \Lambda \bar{ \Lambda } \pi^0$ and $\psip \to \Lambda \bar{ \Lambda } \eta$ are measured to be
$\BR(\psip\to\Lambda\bar{\Lambda}\pi^0)=(1.42 \pm 0.39 \pm 0.59) \times 10^{-6}$ and
$ \BR ( \psip \to \Lambda \bar{ \Lambda } \eta )=(2.34 \pm 0.18 \pm 0.52) \times 10^{-5}$, respectively, where the first uncertainties
are statistical and the second systematic.

Based on the PWA results, it was found that the evident structure around the $\Lambda\eta$ mass threshold could be described by the
$\Lambda(1670)$. The mass and width are determined to be $\mathrm{M}= (1672\pm5\pm6)\ \mathrm{MeV}/c^2$ and
$\Gamma=(38\pm10\pm19)\ \mathrm{MeV}$, which are consistent with those in PDG~\cite{Workman:2022ynf}. The corresponding
product branching fraction is calculated to be
$\BR( \psip \to \Lambda (1670) \bar{ \Lambda } ) \times \BR ( \Lambda (1670) \to \Lambda \eta ) =(1.29 \pm 0.31 \pm 0.62) \times 10^{-5}$,
where the first uncertainty is statistical and the second is systematic.

\begin{table}[htbp]
    \centering
    \caption{Summary of the signal yields and detection efficiencies for each decay mode.}
    \label{tab:value_bf_cal}
    {\begin{tabular}{l |c c }
            \hline
            \hline
            Decay modes                                & $N^{obs}$    & $\epsilon (\%)$ \\
            \hline
            $\psip \to \Lambda \bar{ \Lambda } \pi ^0$ & $23.0\pm6.3$ & 9.0             \\
            $\psip \to \Lambda \bar{ \Lambda } \eta$   & $218\pm17$   & 12.9            \\
            $\psip \to \Lambda(1670)\bar{ \Lambda } $  & $116 \pm 28$ & 12.5            \\
            \hline
        \end{tabular}}
\end{table}

\begin{figure}[htbp]
    \centering
    \includegraphics[width=0.49\textwidth]{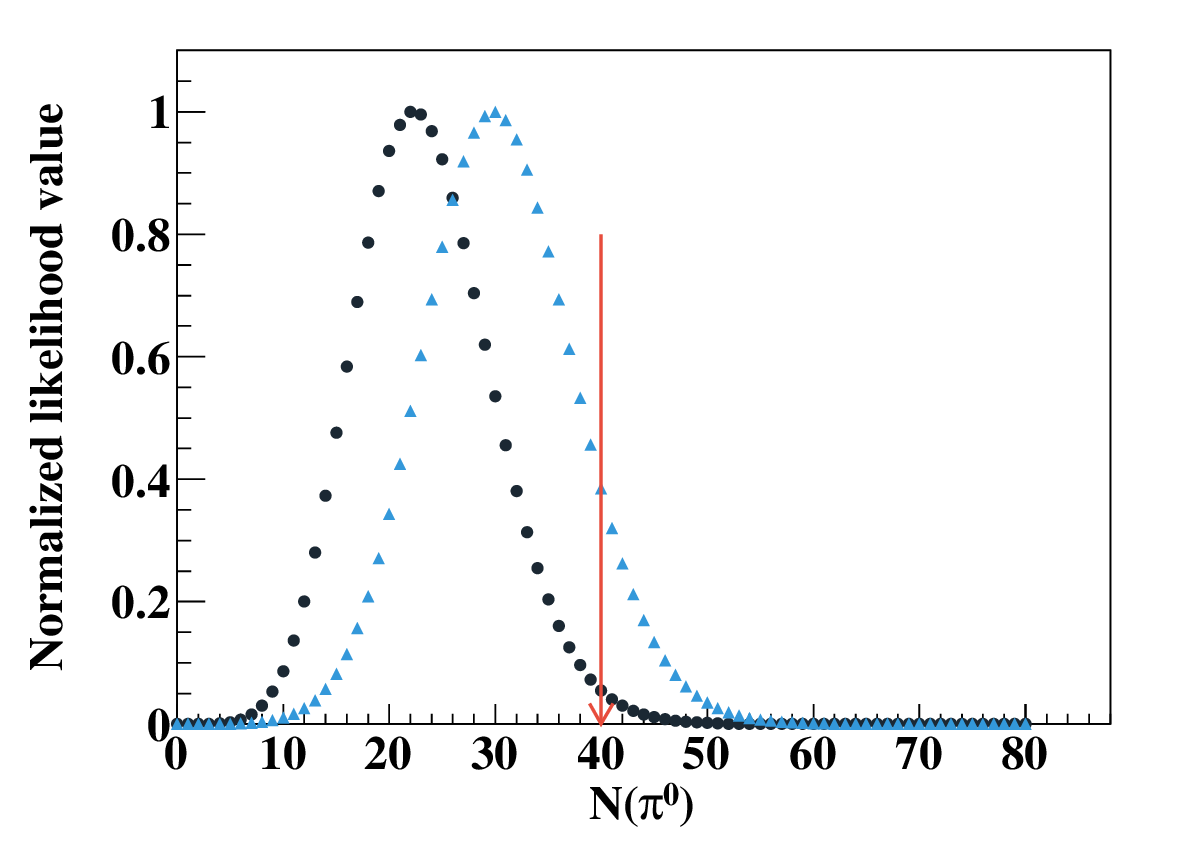}
    \caption{The normalized likelihood distributions. The results obtained with and without incorporating the systematic uncertainties
        are shown in blue dots and black dots, respectively. The arrow is the position of the upper limit on the signal yields at 90\% C.L.}
    \label{fig:Upper_likelihood_pi0}
\end{figure}

Due to the limited statistical significance of the $\psip\to\LLb\pi^0$ signal (3.7$\sigma$), the upper limit of this branching fraction
has been determined. We repeat the maximum-likelihood fits by varying the signal shape, nonpeaking background, peaking background
as well as interference between $\psip$ and continuum amplitudes, and take the most conservative upper limit among different choices.
To incorporate the multiplicative systematic uncertainties in the calculation of the upper limit, the likelihood distribution is smeared
by a Gaussian function with a mean of zero and a width equal to $ \sigma _{ \epsilon }$ as shown below~\cite{Stenson:2006gwf,Liu:2015uha}

\begin{equation}
    L'(n') \propto \int ^{ \infty }_{0} L(n \frac{ \epsilon }{ \epsilon _0}) \exp \left[ \frac{-( \epsilon - \epsilon _0)^2}{2 \sigma ^2_{ \epsilon }} \right] d \epsilon
\end{equation}

\noindent where $L(n)$ is the likelihood distribution as a function of the yield $n$, $ \epsilon _0$ is the detection efficiency and
$ \sigma _{ \epsilon }$ is the multiplicative systematic uncertainty. Figure~\ref{fig:Upper_likelihood_pi0} shows the
likelihood function without and with incorporating the systematic uncertainties. The upper limit on the number of $\psip\to\LLb\pi^0$
events, $N'_{UL}$, is determined to be 40, and the corresponding upper limit of the branching fraction is obtained to be
$\psip\to\Lambda\bar{\Lambda}\pi^0<2.47\times10^{-6}$ at the $90\%$ confidence level (C.L.).

{\centering\section{Summary}}

Using a sample of $448.1 \times 10^6\ \psip $ events collected with BESIII detector at the peak of $\psip$, we performed a study of
$ \psip \to \Lambda \bar{ \Lambda } \pi ^0$ and $ \psip \to \Lambda \bar{ \Lambda } \eta$.

Evidence of the isospin symmetry breaking decay of $ \psip \to \Lambda \bar{ \Lambda } \pi ^0$ is observed with a statistical
significance of $3.7\sigma$, and the corresponding branching fraction is measured to be
$\BR(\psip\to\Lambda\bar{\Lambda}\pi^0)=(1.42\pm0.39\pm0.59)\times10^{-6}$ for the first time. The corresponding upper limit at
the 90\% C.L. is set to be $\mathcal{B}(\psip\to\LLb\pi^0)<2.47\times 10^{-6}$.

In the case of $ \psip \to \Lambda \bar{ \Lambda } \eta $, a PWA is performed to investigate the observed structure around
$\Lambda\eta(\bar{\Lambda}\eta)$. This structure can be described by a $\Lambda(1670)$ with
$\mathrm{M}=(1672\pm5\pm6)\ \mathrm{MeV}/c^2$ and $\Gamma=(38\pm10\pm19)\ \mathrm{MeV}$, which are in good agreement with those
reported by PDG~\cite{Workman:2022ynf}. The corresponding product branching fraction is calculated to be
$\BR( \psip \to \Lambda (1670) \bar{ \Lambda } ) \times \BR ( \Lambda (1670) \to \Lambda \eta )=(1.29 \pm 0.31 \pm 0.62) \times 10^{-5}$.
With the detection efficiency obtained from the weighted MC sample in accordance with the PWA results, the branching fraction of
$ \psip \to \Lambda \bar{ \Lambda } \eta $ is measured to be
$ \BR ( \psip \to \Lambda \bar{ \Lambda } \eta )=(2.34 \pm 0.18 \pm 0.52) \times 10^{-5}$.

Compared with the branching fraction of $ \jpsi \to \Lambda \bar{ \Lambda } \pi ^0$ and $ \jpsi \to \Lambda \bar{ \Lambda } \eta $
\cite{BESIII:2012jve}, the ratio between the branching fractions of $\psip$ and $\jpsi$ decaying to the same hadronic final state
is defined as $\mathcal{Q}_h$. Taking PHSP factors into account, the ratio of $\LLb\pi^0$ and $\LLb\eta$, $\mathcal{Q}_h(\pi^0)$ and
$\mathcal{Q}_h(\eta)$, are calculated to be $(1.4 \pm 0.7)\% $ and $(2.3 \pm 0.6)\% $, respectively, both contradicting the 12\% rule
significantly. \\

\acknowledgments

The BESIII collaboration thanks the staff of BEPCII and the IHEP computing center for their strong support. This work is supported in
part by National Key R\&D Program of China under Contracts Nos. 2020YFA0406300, 2020YFA0406400; the Chinese Academy of Sciences (CAS)
Large-Scale Scientific Facility Program; Joint Large-Scale Scientific Facility Funds of the NSFC and CAS under Contract No. U2032110;
National Natural Science Foundation of China (NSFC) under Contracts Nos. 11635010, 11735014, 11835012, 11935015, 11935016, 11935018,
11961141012, 12022510, 12025502, 12035009, 12035013, 12192260, 12192261, 12192262, 12192263, 12192264, 12192265; CAS Key Research Program
of Frontier Sciences under Contract No. QYZDJ-SSW-SLH040; 100 Talents Program of CAS; INPAC and Shanghai Key Laboratory for Particle
Physics and Cosmology; ERC under Contract No. 758462; European Union's Horizon 2020 research and innovation programme under Marie
Sklodowska-Curie grant agreement under Contract No. 894790; German Research Foundation DFG under Contracts Nos. 443159800, Collaborative
Research Center CRC 1044, GRK 2149; Istituto Nazionale di Fisica Nucleare, Italy; Ministry of Development of Turkey under Contract No.
DPT2006K-120470; National Science and Technology fund; STFC (United Kingdom); The Royal Society, UK under Contracts Nos. DH140054,
DH160214; The Swedish Research Council; U. S. Department of Energy under Contract No. DE-FG02-05ER41374

\bibliography{ref_V3.7}

\end{document}